# Determinants of ICT Adoption Among Small Scale Agribusiness Enterprises In Somalia

Husein Osman Abdullahi[1], Abdikarim Abi Hassan[2], Murni Mahmud[3], Abdifatah Farah Ali[4]

[1,4]Faculty of Computing, [2]Faculty of Engineering, SIMAD University, Somalia
[3]Kulliyah of ICT, International Islamic University Malaysia (IIUM), Malaysia

[1]husein@simad.edu.so , [2]abdulkarimabi@simad.edu.so , [3]murni@iium.edu.my

*Abstract -* The use of Information and Communication Technology (ICT) can advance the Agricultural Business sector, particularly in a country seeking opportunities to explore the sector. There is evidence that ICT has made significant contributions to agribusiness because it allows enterprises to manage their operations, and it has major impacts on the business. However, the critical factors that motivate the adoption of new innovative technology by agribusiness enterprises are underexplored. The literature has indicated ICT adoption among small-scale agribusiness enterprises in Somalia is not fully understood. Nevertheless, this study addresses this gap by investigating the adoption of ICT among small-scale agribusiness enterprises in Somalia. The paper reports the use of the Technology, Organization, Environment (TOE) framework. An online survey has been conducted with random sampling for data collection, with 107 respondents.  The respondents are from agribusiness staff and farmers from various agricultural companies in Somalia. After quantitative data analysis, the results indicated that relative advantage, complexity, top management support, and competitive pressure factors are significant contributors to ICT adoption in Somalian agribusiness enterprises, while ICT costs and vendor support are not significantly related to the adoption of ICT in agricultural business. This study concludes that ICT adoption in Somalia is inspired by insight and motivation rather than financial and external support. Understanding these factors leads to a better understanding of ICT adoption in Somalia. Additionally, it enriches the literature about the agriculture business on the African continent

## I. INTRODUCTION

Agriculture acts as a backbone to the economy of developing countries, playing a significant role in economic growth (Byerlee, De Janvry, et al., 2009; Kante, Oboko, et al., 2017). It increases annual income and has a positive impact on the local market and businesses. Agricultural production has contributed to yearly income and has a positive impact on local market access to food supply which contributes to agricultural sustainability through improved agricultural best practices (OECD-FAO, 2015).  In Africa, agricultural production has increased gradually, and its value has almost tripled (+160%), which is almost equal to South America, but less than Asia in terms of the growth and sustainability of agricultural production (Blein, 2013).

In developed countries, the development and advancement in agricultural products have been largely attributed to the latest technologies. At the same time, developing countries, e.g., in Africa, have been rated with low agricultural produce (Aker, 2011; Aleke, Ojiako, et al., 2011, Kante, Oboko, et al., 2017). This might be caused by a lack of infrastructure for the dissemination of information about crucial subjects such as farm input, productivity, efficient farming skills, etc. These countries also have limited access to recent technological advances directly or indirectly related to farming and its practices (Kante, Oboko, et al. 2019). Other challenges include poor financing, planning, and unfair competition in the agricultural market (Chavula 2014).

Sufficient access to the latest ICT developments in the agriculture sector contributes significantly to agricultural sustainability, providing small-scale agribusiness with a competitive advantage in the new economy (Chavula 2014, Hoque, Saif, et al. 2016, Aldosari, Al Shunaifi, et al. 2017). Thus, the importance of information technology in agribusiness enterprises cannot be over-emphasized. However, African agribusiness enterprises have limited access to advanced technology, particularly in Somalia.

The Somalian GDP increased by 3.1% in 2018, driven by increased agricultural production, private investment , and donor inflows (World Bank 2019).
The Somali agricultural sector contributes about 60 percent to the country's gross domestic product (GDP), according to the World Bank report. According to Africa Agriculture Status Report (2018), the success of agribusiness depends on the investment of information access and communication technologies.

Previous studies have focused on ICT adoption in small-scale agribusiness in other countries.  Aleke *et al.* (2011) conducted a study of ICT adoption in small-scale agribusiness in Nigeria. Burke (2010) investigated the adoption and use of ICT in small-scale agribusiness firms in Hawaii, United States of America. However, there is not much evidence of research on ICT adoption in small-scale agribusiness in Somalia. Therefore, the state of agribusiness in Somalia is not yet well understood.

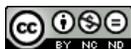

The purpose of this research is to determine the factors influencing ICT adoption in Somalia's small-scale agribusiness. This paper focuses on two areas: firstly, investigating current ICT usage in Agribusiness enterprises and, secondly, identifying the factors that affect the adoption of ICT services in Somalia. The study has been conducted in all parts of Somalia, where agriculture and livestock are the main economic sources in the region.

In explaining the research, this paper is divided into four sections: Section I presents the Literature review and TOE framework; section II also develops the material and methodologies of the research; section III presents the Research model and hypotheses. The results and discussion of the research are presented in section IV, which also concludes with limitations and directions for future research.

## II. LITERATURE REVIEW

The impact of ICT on agriculture in 34 African countries for 11 years was analyzed by Chavula (2014), using unbalanced panel data regression analysis. The results indicate that ICT helped to increase agricultural production in these countries. It is noteworthy that the Internet was not found to have been used to its critical potential. Rather most of the impact was observed to be from communication channels such as phones (Chavula 2014).

Another study also discusses the application, effect, and use of ICT in small and medium-sized companies. Top management support, government support, financial support, and recognition of benefits have been considered as the main factors in these companies' adoption (Hoque, Saif, et al., 2016). The research found a clear positive association between financial support and the implementation of ICT and a negative association with government support. (Hoque, Saif, et al. 2016).

A study looked into factors impacting the adoption of information technology in a sample of cereal farmers (Kante, M. et al., 2019). The farmers have been provided necessary ICT support for the dissemination of vital information regarding farm practices of their niche. The adoption was affected by factors such as compatibility, simplicity, the relatedness of information, and observability to the farmers. Social pressure was reported to have almost minimal effect on adoption (Kante, Oboko, et al. 2019).

Qualitative studies were also carried out to examine the trends of IT influence on the agriculture context. For instance, Botsiou and Dagdilelis (2013) found that many families in Greek farming communities contributed to the use of ICT by providing relevant information and were considered to be key drivers for agricultural communities. Adoption of and opposition to ICT was found to be largely dependent on the awareness of individuals. It was concluded that digitally educated people could be major drivers of the adoption of ICT in such communities (Botsiou and Dagdilelis, 2013).

ICT adoption has been studied mostly as being static, relying on conceptualizing factors as variables by focusing on a single decision outcome. These approaches have less consideration of the dynamic nature of variables and the effect of various challenges during the adoption process. Using a hybrid approach, study six (6) examines fourteen (14) key factors influencing ICT adoption, studying the effect of changes and challenges at various levels on adoption in SMEs'. Several factors, such as ease of use, managerial time and customer focus, etc., have a visible impact at all stages, while others might impact only at certain stages. However, all of them are dynamic and influence adoption across multiple stages of the adoption process.

The framework presented serves as an analytical tool for pre and post-adoption, focusing on the nature of factors through multiple stages. Information dissemination is a key factor for increased agricultural yields and provides innovative methods of management in developing countries. Various countries have tried multiple programs for ICT intervention and adoption, but factors such as scale, sustainability, relevance, and responsiveness are impeding factors for their success (Eze, Chinedu-Eze, et al., 2019).

Aker (2011) highlights the importance of conducting rigorous training programs continually, customized to the needs of the specific communities. The impact needs to be measured by disseminating relevant information over a relevant channel in order to overcome the problem of responsiveness (Aker 2011).

ICT is a critical factor in bridging the rapidly growing divide between rural and urban economies. Many rural economies are largely dependent on agricultural produces. Efforts such as having widespread Internet access in these communities have been implemented, but users are unable to take advantage. On an individual level, factors such as age, education level, access to a computer, and internet skill affect the adoption. Most importantly, it has been found that pieces of training and perceived usefulness are major factors aiding adoption (Moghaddam and Khatoon-Abadi 2013).

Similar findings were authored by Aldosari (2017) in reference to farming communities in Pakistan. Young farmers are more willing to accept ICT intervention. Similarly, educated farmers accept and adopt new innovations readily. (Aldosari, Al Shunaifi, et al., 2017).

## III. MATERIALS AND METHODS

This research utilizes the framework for Technology, organization, Environment (TOE), and it also employs a quantitative data analysis through an online survey via random sampling of one hundred and seven (107) respondents whom are agribusiness staff and farmers from various agricultural companies in Somalia.

### A. Materials and Methods

The theory behind this research is rooted in Tornatzky and Fleischer's (1990) model of Technological Innovations, which approaches innovation by considering the technology and organizational environment. The TOE model, a multifaceted corporate innovation method, discusses how the technical, organizational, and environmental dynamics affect innovation adoption and execution (Tornatzky & Fleischer, 1990).

The inclusion of these dynamics makes the model much suited to the current research to perform a study on the adoption of agribusiness. The technological context refers to relevant internal and external innovations for the underlying domain. These include both innovations currently being utilized as well as other available ones whose potential has not been exploited yet. The organizational dimension focuses on sector features and resources, looking at size, products, services, employees, etc., while the environment relates to the climate of organizational operations. It tends to study interrelated business dimensions, competition, regulations, and support partners at the same time, providing an overall operational view of the market (Tornatzky, Fleischer et al. 1990, Zhu and Kraemer 2005).

The research model contains seven-factor, each of which was measured with multiple items. The items were adopted from the extant literature to preserve content validity (Straub, Boudreau, et al. 2004). The theoretical basis for each factor is described in the subsequent section, and Figure 1 shows the research model.

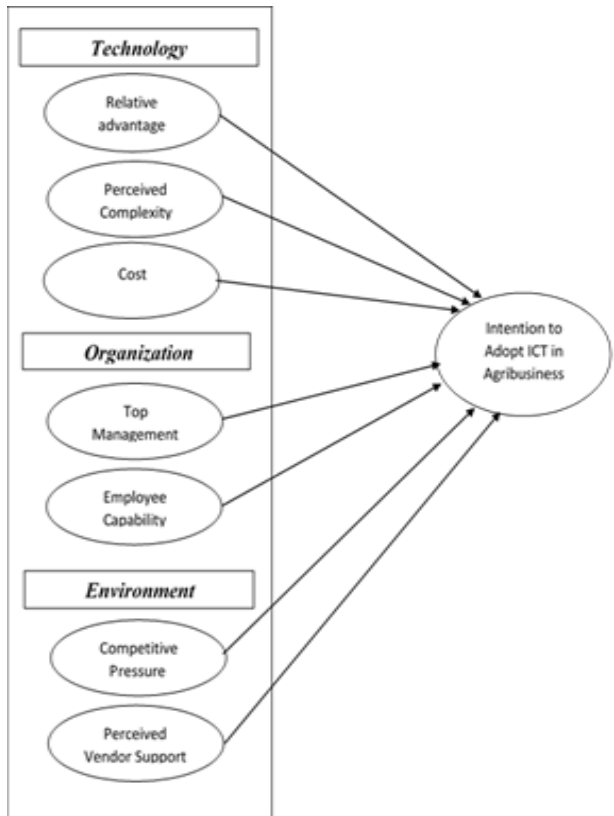

**Figure 1 Technology, Organisation and Environment (TOE) model for ICT Adoption in Agribusiness**

*B. Research Hypothesis*
**TECHNOLOGY**
In the research model, Technology is scrutinized by using three components: Cost, Perceived complexity, and Relative Advantage. Each item is described respectively. *Relative Advantage* is said to be the level to which the annovation is "the degree to which an innovation is seen as more significant than the idea it represents" (Rogers, 2003, p. 15).

The positive perceptions of the benefits of ICT and connectivity can motivate organizations to adopt new and innovative technologies (Thong 1999). For organizational growth and development, factors relating to competition, quality, and social status are commonly regarded as the essential aspects of relative gain (To and Ngai 2006).

Moreover, companies only adopt new innovative technology when they feel that this technology can provide opportunities to enhance their businesses and solve current problems (Premkumar and Roberts, 1999). Therefore, the relative advantage is a determinant of ICT adoption in agribusiness in Somalia. So, the hypothesis below is put forward

**H1:** Relative Advantage has a positive influence on ICT adoption in agribusiness

Rogers (1983) found that *Perceived Complexity* affects technology adoption. Technology is viewed to be more convenient to use, and the process of adopting is much faster. Cooper and Zmud (1990) found that complex technologies need more technical skills, organizational efforts, and strong implementation to increase their chances of adoption in agricultural business. This gives rise to the formation of the next hypothesis of this research:

**H2.** The adoption of ICT among Somali agribusiness has a positive relation with Perceived Complexity

The *Cost* incurred for ICT implementation is another important factor affecting adoption. Past studies illustrate that higher investment costs impact adoption negatively. (James 2003, Yeung, Shim, et al. 2003). As presented by a report of OECD (2000), the cost is perceived to be a significant issue for owners and managers of agribusiness. Thus, the following hypothesis is suggested:

**H3.** The adoption of ICT among small scale agribusinesses is negatively related to ICT cost

**ORGANISATION**
Innovation adoption requires qualified staff to handle the technological change being implemented. Usually, *small and medium organizations* lack internal technical specialist staff. External contracting is too expensive and thus unaffordable. This poses a serious threat to ICT innovation in any environment. The literature highlights a shortage of adequately well-trained internal technological innovation experts in agricultural-related companies. This largely affects the maturity of ICT in agribusiness firms (Caldeira and Ward 2003, Fisher and Howell 2004, Ghobakhloo, Hong, et al. 2012). A significant number of specialist staff is expected to motivate the behavioral intention of agribusinesses to use ICT and other technological innovations. This leads to the following hypothesis:

**H4.** Employee Capability has a significantly positive impact on an organization's ICT Adoption

*Top Management Support* refers to senior manager policies devoted to promoting the acceptance of upcoming changes (Hult, Ketchen, et al. 2007). The highest rate of acceptance and use of innovative technologies is typically determined by the management's decisions (Rogers 2003). Moreover, managers with a strong desire to adopt new technological innovations can have a greater impact on their business decisions. Thus, their support is very significant in influencing ICT adoptions and diffusions. Top management support has been proven as a motivating factor for other organizational leaders to act appropriately to support innovation. Most recent research has found that the level of technology acceptance in organizations progresses positively because of top management support (Chatterjee, Grewal, et al. 2002, Martins, Oliveira, et al. 2016).
Top management support and involvement in ICT implementation and adoption in agribusiness are very crucial. Managers can positively influence the adoption of ICT in agribusiness by offering support at the company's highest level. Thus, the positive influence hypothesis is proposed as follows:

**H5.** Top Management Support will positively affect the adoption of ICT in agribusiness

**ENVIRONMENT**

The environment is an important construct in the research model. 2 items are used for measurement: Competitive Pressure and Perceived Vendor Support. Competitive Pressure refers to the degree of pressure a firm feels in the same industry from its competitors (To and Ngai 2006). Moreover, Competitive Pressure means the external pressures that force companies to change and adapt information and communications technology (Obal 2017). In a specific industry, early adopters of technology typically have a first-mover advantage. In addition to that, competitors will also consider the implementations of ICT platforms in agribusiness to manipulate customers in the market places which are open and fair market opportunities.

When their rivals are the first to embrace advances in ICT, the resulting higher pressure will motivate companies to track their lead. Organizations tend to look to embrace the most recent ICT to improve their competitive advantage against the challenge of furious competition (Zhu, Kraemer, et al. 2006).

The current commerce climate is both volatile and intense, with businesses in the same sector being forced to mimic industry pioneers' behavior and achieve competition due to increased market pressures (Lin and Lin 2008). In a nutshell, organizations can choose to use ICT as a new strategy to outperform their rivals in order to hold a desirable market position. Thus, the hypothesis below is suggested:

**H6.** Competitive Pressure will positively affect a firm's intention to adopt ICT.

Next, supplier efforts and Vendor Support is "the availability of support for implementing and using an information system" (Premkumar and Roberts 1999). Furthermore, organizations may be more willing to try new technology if they think that there is enough support from the vendors. Past studies (Hultink, Griffin, et al. 1997, Frambach, Barkema, et al. 1998, Woodside, Biemans, et al. 2005) have attempted to tie the marketing efforts of suppliers with the adoption decisions of the client. According to these studies, there is a link between the marketing efforts of the supplier and the adoption decision of the client (Park and Chen 2007), with providers functioning as both a source of information and reassurance about their capabilities (Weigelt and Sarkar 2009). Therefore, the following hypothesis is offered:

**H7.** There is a significant relationship between Perceived Vendor support and ICT adoption in Agribusiness.

## IV. RESULTS AND DISCUSSION

Table 1 presents the demographic data, highlighting a ratio of 92 males to 15 females. Respondents were categorized into five age groups: 18-25 years (n=48), 26-35 years (n= 51) 36-40 years (n=3), 41-49 years (n=2), 50 years and above (n=3). The education level of respondents was categorized into four groups: Diploma (n=2), Bachelor (n=73), Master (n=29), And Ph.D. (n=3). Observations of the demographics indicate more than 80% of the Somalis that are involved in agribusiness are male and aged between 18-35 years (92.5%). They may be considered young and energetic. The respondents are also highly educated; education may have made them open and aware to explore new opportunities and challenges.

**Table 1: Demographics of the study**

| Distribution | | Frequency | Percentage (%) |
|---|---|---|---|
| **Gender** | Male | 92 | 85.9% |
| | Female | 15 | 14.1% |
| | | 107 | 100% |
| **Age** | 18-25 years | 48 | 44.8% |

| | 26-35 years | 51 | 47.7% |
|---|---|---|---|
| | 36-40 years | 3 | 2.8% |
| | 41-49 years | 2 | 1.9 |
| | >50 | 3 | 2.8% |
| | | 107 | 100% |
| **Education Level** | Diploma | 2 | 1.9% |
| | Bachelor | 73 | 68.2% |
| | Master | 29 | 27.1% |
| | PHD | 3 | 2.8% |
| | | 107 | 100% |

**Measurement Model**

To evaluate the measurement model, convergent validity and discriminant validity were examined.

**Convergent Validity**

According to Hair Jr, Hult et al. (2016), convergent validity is a measure of internal consistency that assesses the suitability of correlation of items within a scale to measure the same construct (Hair, Ringle, et al. 2011). They are determined through factor loading, Average Variance Extraction (AVE), Cronbach's Alpha, and Composite Reliability (CR) measures of Dijkstra-Henseler's rho (ρA) and Jöreskog's rho (ρc). This study found that the item loading was higher than 0.7, the AVE was above 0.5, and the value of Dijkstra-Henseler's rho (ρA) and Jöreskog's rho (ρc) was more than 0.7, as shown in Table 2. These results validate that all criteria were satisfactory as the three criteria meet the recommended threshold values. At the same time, one item was dropped due to low factor loading

**Table 2: Convergent Validity**

| Construct | Items | Loading | Composite reliability (CR) | The average variance extracted (AVE) |
|---|---|---|---|---|
| **ICT adoption in agribusiness** | ICTA1 | 0.918 | 0.93 | 0.816 |
| | ICTA2 | 0.902 | | |
| | ICTA3 | 0.889 | | |
| **Perceived Complexity** | PC1 | 0.843 | 0.89 | 0.73 |
| | PC2 | 0.849 | | |
| | PC3 | 0.871 | | |
| **Perceived Cost** | PCO1 | 0.554 | 0.759 | 0.629 |
| | PCO2 | 0.975 | | |
| **Perceived Competitive Pressure** | PCUP1 | 0.847 | 0.916 | 0.732 |
| | PCUP2 | 0.892 | | |
| | PCUP3 | 0.805 | | |
| | PCUP4 | 0.878 | | |
| **Perceived Employee Capability** | PEC1 | 0.867 | 0.933 | 0.776 |
| | PEC2 | 0.86 | | |
| | PEC3 | 0.909 | | |
| | PEC4 | 0.885 | | |
| **Perceived Relative Advantage** | PRA1 | 0.844 | 0.915 | 0.685 |
| | PRA2 | 0.889 | | |
| | PRA3 | 0.859 | | |
| | PRA4 | 0.852 | | |
| | PRA5 | 0.677 | | |
| **Perceived Vendor Support** | PVS1 | 0.904 | 0.93 | 0.815 |
| | PVS2 | 0.903 | | |
| | PVS3 | 0.902 | | |
| **Top Management Support** | TMS1 | 0.876 | 0.911 | 0.72 |
| | TMS2 | 0.87 | | |
| | TMS3 | 0.851 | | |
| | TMS4 | 0.794 | | |

**Discriminant Validity**

The Fornell-Larcker (1981) criterion has been criticized due to its lack of reliability to detect the lack of discriminant validity in common research situations (Henseler, Ringle, et al. 2015). Therefore, this study employed a heterotrait-monotrait ratio of correlations (HTMT) based on the multitrait-multimethod matrix. (Henseler, Ringle, et al., 2015). The study assessed the HTMT value, whether the value is greater than the $HTMT_{.85}$ value of 0.85 (Kline 2015), or the $HTMT_{.90}$ value of 0.90 (Gold, Malhotra, et al. 2001) because if the HTMT is greater than the aforementioned values, then discriminant validity is questionable. All HTMT values passed the $HTMT_{.90}$ and $HTMT_{.85}$ threshold test as shown in Table 3, indicating that discriminant validity has been established.

**Table 3: Discriminant Validity Heterotrait-Monotrait (HTMT) Criterion**

| | ICT Adoption | PC | PCO | PCUP | PEC | PRA | PVS |
|---|---|---|---|---|---|---|---|
| ICT Adoption | | | | | | | |
| PC | **0.474** | | | | | | |
| PCO | 0.108 | **0.148** | | | | | |
| PCUP | 0.827 | 0.345 | **.070** | | | | |
| PEC | 0.804 | 0.348 | .109 | **0.811** | | | |

| | | 0.380 | .065 | 0.777 | **0.790** | | |
|---|---|---|---|---|---|---|---|
| PRA | 0.829 | | | | | | |
| PVS | 0.655 | 0.402 | .074 | 0.737 | 0.807 | **0.677** | |
| TMS | 0.855 | 0.407 | .072 | 0.823 | 0.757 | 0.849 | **0.702** |

**Structural Model**

This study utilizes a Partial Least Square (PLS) regression, which is an extension of the classical multiple linear regression model. According to Hair Jr et al. (2016), R-squared, standard beta, t-values via a bootstrapping procedure with a resample of 5000 and an effect size ($f^2$), and the predictive relevance ($Q^2$) have to be examined to assess the structural model. In this study, all of these matrices and parameters have been assessed, as shown in Table 4.

| **Table 4: Structural Model-Hypothesis testing** | | | | | | |
|---|---|---|---|---|---|---|
| HS | Path relationship | Std. Beta | SE | t-value | P Values | Decision |
| H1 | PC -> ICT Adoption | 0.117 | 0.059 | 1.978 | 0.024*** | Supported |
| H2 | PCO -> ICT Adoption | -0.055 | 0.094 | 0.581 | 0.281 na | Not Supported |
| H3 | PCUP -> ICT Adoption | 0.228 | 0.113 | 2.016 | 0.022*** | Supported |
| H4 | PEC -> ICT Adoption | 0.243 | 0.109 | 2.229 | 0.013 *** | Supported |
| H5 | PRA -> ICT Adoption | 0.206 | 0.113 | 1.818 | 0.035*** | Supported |
| H6 | PVS -> ICT Adoption | -0.081 | 0.098 | 0.824 | 0.205 na | Not Supported |
| H7 | TMS -> ICT Adoption | 0.281 | 0.104 | 2.697 | 0.004*** | Supported |
| *p* <0.05*, *p*<0.01**, *p*<0.001***, na= not significant | | | | | | |

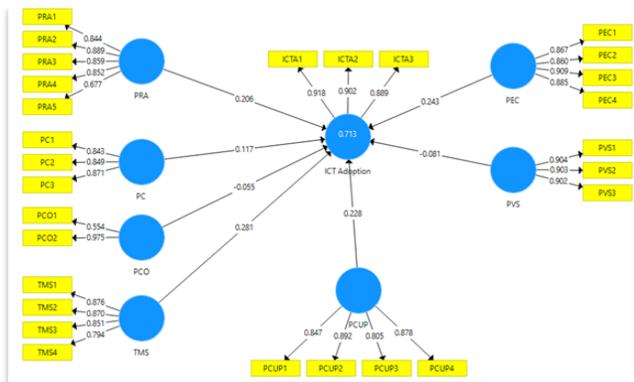

**Figure 2: Structural Model**

After analysis, the study found that Perceived Complexity (PC) has a significant relationship with ICT adoption in small-scale agribusiness (β = 0.117, p < 0.024). Therefore, H1 is supported. Also, Perceived Competitive Pressure (PCUP) has a significant relationship with ICT adoption in small scale agribusiness (β = 0.228, p < 0.022). Therefore, H3 is supported. Furthermore, Perceived Employee Capability (PEC) was found to have a significant relationship with ICT adoption in small-scale agribusiness (β =**0.243**, p < 0.013), and therefore, H4 is supported. Also, Perceived Relative Advantage (PRA) has a significant relationship with ICT adoption in small-scale agribusiness (β =**0.206**, p < **0.035**), hence H5 is supported. Finally, Top Management Support (TMS) has a significant relationship with ICT adoption in small-scale agribusiness (β =**0.281**, p < **0.004**), therefore, H7 is supported.

In contrast, the results point out that Perceived Cost (PCO) (β = **-0.055**, p> 0.281) and Perceived Vendor Support (PVS) (β = **-0.081**, p> 0.205) had no impact on ICT Adoption. As a result, H2 and H6 were not supported. The value of the R-squared indicates 0.713, as shown in Table 4, which indicates that all the 9 predictors together can explain 71.3% of the variance in ICT adoption in small-scale agribusiness.

**V. Discussion**

The paper has discussed the determinants of ICT adoption by agribusiness companies in Somalia. In achieving its objectives, a multi-perspective approach was implemented by considering variables in the organizational, technological, and environmental aspects for developing the theoretical framework. A field survey has been conducted and responded to by selected stakeholders, employees, and decision-makers of agribusinesses. They also confirmed the proposed research model. The dimensions of the research are elaborated on in the following sections.

**TOE Framework**

The TOE framework has served as a guide for the research to identify the significant determinants for the success of ICT Adoption in agribusiness in Somalia. Each factor will be discussed below.

**Technology**

Of the three technological factors that were hypothesized as determinants of the technology adoption in Somalian Agribusiness enterprise, the results indicate that relative advantage and complexity are significant contributors to ICT adoption. Comparatively, ICT costs were found to have significantly less impact on adoption, which could be contributed by the willingness of each participant to cover the cost. It could also be attributed to funding availability from the government or other agencies.

The significance of relative advantage is consistent with previous studies (Premkumar and Roberts 1999, Gamal Aboelmaged 2010, Wang, Wang et al. 2010, Maduku, Mpinganjira, et al. 2016). In this study, relative advantage indicates that respondents have a positive outlook about the advantages of ICTs in agribusiness. Therefore, they are more likely to have a positive intention to accept ICT within their firms. This result indicates that expected relative advantage is an important determinant of the intention of agribusiness to implement ICT innovations.

Additionally, the results indicate that the complexity of technology is an insignificant factor, which means there is no positive relation to adopting ICT in agricultural business in Somalia. This is also consistent with past studies, which conclude that the complexity of technology is not a critical factor in the adoption of cloud computing (Low and Chen, 2011; Lin and Chen, 2012). However, one other study contradicts this finding and suggests that external technical support for ICT vendors was a key important factor of ICT adoption among Singaporean small businesses. (Thong, 2001)

The findings also indicate a significant negative relationship between perceived cost and behavioral intention to use ICT platforms in agribusiness enterprises in Somalia. This may suggest that the respondents believe that the higher the cost of ICT adoption in agribusiness, the less likely it to be accepted. This result is supported by other similar studies (Madlberger 2009, Ramayah, Ling, et al. 2016) that have found a negative relation between technological adoptions and perceived cost. However, this poses a challenge for adopting ICT platforms in agribusiness firms. IT experts and decisions makers could address this obstacle by emphasizing the importance of technology for the successful operation of companies. On the other hand, when potential adopters realize the benefits of adopting ICT tools, then the costs may no longer act as a barrier to adoption.

**Organization**

The current study reveals that the ability of employees to adopt ICT among small-scale agribusiness enterprises in Somalia is another important factor. Literature shows that this result contradicts other earlier research (Ruivo, Oliveira, et al. 2014), (Ramayah, Ling, et al. 2016). It was found that the IT knowledge and expertise of employees is not a significant factor for their intention to act. Nevertheless, it is rational to purport that a great deal of employee trust in their ability to handle the complexities of ICT use would be linked with high probability of intention of agribusinesses to use ICT tools.

Previous studies emphasized that a lower degree of employee willingness to use software will be an obstacle to the behavioral intention of businesses to apply technology (Ghobakhloo, Hong, et al. 2012)

Top management support (TMS) has also been significant. This result goes well with former studies on such technologies as e-procurement, cloud computing, E-commerce, Enterprise Systems, and Electronic Data Interchange (EDI) (Premkumar and Roberts 1999, Ramdani and Kawalek 2007, Teo, Lin, et al. 2009, Borgman, Bahli, et al. 2013, Ruivo, Oliveira, et al. 2014). Ruivo, Oliveira et al. (2014) assumed that top management in the enterprise could exert their influence on the adoption of cloud computing by using money as well as other resources. TMS is considered a very crucial determinant in the adoption of ICT. Allocation of resources and business process re-structuring may be required as a result of technological acceptance. Hence, the support of top management is needed in the process of ICT adoption.

**Environment**

This study showed a significant relationship between competitive pressure and the adoption of ICT in agribusiness enterprises. The results from previous studies signify that competitive pressure is a vital component influencing companies' adoption of innovative technologies (Ghobakhloo, Arias-Aranda, et al., 2011). As indicated by Chwelos, Benbasat et al. (2001), (Wang and Cheung 2004, Lin 2014), the relationship between competitive pressure and technology adoption at a company level is important. This implies that there is a positive intention amongst small and medium-sized enterprises to adopt ICT.

On the other hand, the effect of vendor support accessibility on the plan to implement ICT in agribusiness has been found to be insignificant. However, this finding is inconsistent with Al-Qirim (2007), who suggested that the availability of vendor support in IT technology can result in agribusiness companies adopting innovative technology.

### IV. Conclusion

ICT plays an important role in agricultural business in helping to improve productivity, disseminating information to the farmers, and supporting agri-businesses. It is notable that the challenges of ICT in agriculture relate to implementation. This can be overcome by framing a good strategy for the ICT adoption process in agricultural companies.

The main contribution of this study is that it provides an empirical investigation of TOE factors that are found to have influenced ICT adoption in small-scale agribusiness enterprises in Somalia. The study has tried to introduce and examine a model that hypothesizes the effects of technology, organizational and environmental aspects on ICT adoption within agricultural business companies. In addition, random sampling for data collection was carried out in an online survey. The number of respondents was 107. The respondents were farm employees and farmers from different agricultural companies in Somalia.

The findings of this research can benefit the staff and managers of agribusiness companies by providing them with a better understanding of what factors can influence the adoption of ICT in agricultural business enterprises. In particular, the study makes a valuable contribution to the field of agricultural business in Somalia which often has little coverage in the literature. The results suggest that relative advantage, complexity, top management support, and competitive pressure have positive effects on the adoption of ICT use in agribusiness. Other factors such as cost and vendor support do not have a significant influence on the adoption of ICT platforms in agribusiness companies.

The limitation of this study lies in the research scope. It focused only on the intended factors for ICT adoption in small-scale agribusiness in Somalia. It is very challenging to make comparisons in the context of Somalia because there is a lack of literature on the subject

However, this study enriches the literature on ICT in agricultural business in Somalia, which can be leveraged to further study of the African continent at large. Future research may explore possibilities to improve ICT infrastructure and applications, ICT policy in agriculture, and implementing advanced technologies. Through these findings, we can be optimistic that ICT access can be made accessible to most farmers and staff in agribusiness in rural and urban places.

It is hoped that agri-business managers and administration will strategically provide and improve staff capacity. It is believed that Government support for agribusiness will accelerate its growth by creating opportunities and encouraging the use of ICT in the field. Ultimately, there is a need for governments to increase their commitment to productive investments such as technology, agricultural research centers, and rural infrastructure for local rural communities.